\documentclass[preprint,12pt]{elsarticle}
\usepackage{amssymb}
\usepackage{amsmath}
\usepackage{array}
\usepackage{caption}
\usepackage{makecell}
\usepackage{makebox}
\usepackage{xcolor}
\usepackage{colortbl}
\usepackage{graphicx}
\usepackage{float}

\makeatletter
\def\ps@pprintTitle{%
 \let\@oddhead\@empty
 \let\@evenhead\@empty
 \def\@oddfoot{}%
 \let\@evenfoot\@oddfoot}
\makeatother

\begin{document}
\begin{frontmatter}

\title{Asymmetric Iterated Prisoner's Dilemma on BA Scale-Free Network}

\author[a]{Yunhao Ding\fnref{fn1}}
\fntext[fn1]{First author}

\author[b]{Chunyan Zhang\fnref{fn2}}
\fntext[fn2]{Second author}

\author[b]{Jianlei Zhang\corref{cor1}}
\ead{jianleizhang@nankai.edu.cn}
\cortext[cor1]{Corresponding author}

\affiliation[a]{organization={Department of Automation, College of Artificial       
            Intelligence},
            addressline={Nankai University}, 
            city={Tianjin},
            postcode={300071}, 
            country={China}}

\affiliation[b]{organization={Tianjin Key Labortory of Intelligence         
            Robotics},
            addressline={Nankai University}, 
            city={Tianjin},
            postcode={300071}, 
            country={China}}

\begin{abstract}
In real-world scenarios, individuals often cooperate for mutual benefit. However, differences in wealth, reputation, and rationality can lead to varying outcomes for similar actions. Besides, in complex social networks, an individual's choices are frequently influenced by their neighbors. To explore the evolution of strategies in realistic settings, we conducted repeated asymmetric prisoner’s dilemma experiments on a weighted Barabási-Albert (BA) scale-free network using a memory-one strategy framework. First, our analysis highlighted how the four components of memory-one strategies affect win rates. Second, during strategy evolution on the network, two key strategies emerged: "self-bad, partner-worse" and "altruist". Finally, by introducing optimization mechanisms, we increased the cooperation levels among individuals within the group. These findings offer practical insights for addressing real-world problems.
\end{abstract}


\begin{highlights}
\item We analyze and compare the characters of each component in the framework of memory-one strategy.
\item We find "altruists" strategy and "self-bad, partner-worse" strategy within an iterated asymmetric prisoner's dilemma game on weighted BA scale-free network.
\item We explore methods to enhance the average fitness of the population.
\end{highlights}


\end{frontmatter}



\section{Introduction}
\label{sec1}

Cooperation refers to the behavior where individuals coordinate to achieve better outcomes driven by common interests\cite{axelrod1985achieving}. In the biological realm, cooperative behaviors are ubiquitous, ranging from foraging activities among animals to relations between nations\cite{cheng2023evolution}. To study the impact of individuals' choices to cooperate or not under complex conditions on the benefits to both parties, game theory and evolutionary game theory have emerged successively\cite{sigmund2011introduction}. Game theory provides the mathematical framework for analyzing scenarios characterized by conflict or competition. Evolutionary game theory, a branch of game theory, integrates concepts from evolutionary biology to explore strategic choices within a population and the dynamic processes of behavioral evolution\cite{weibull1997evolutionary, zhao2023mechanisms}.

The Prisoner's Dilemma (PD) is a classic model in game theory, originating from a scenario involving two captured prisoners who are unable to communicate with each other. PD presents a seemingly paradoxical problem: when faced with the choice between betrayal and cooperation, the rational choice for each prisoner is to betray, because, regardless of the other’s decision, confessing yields the best individual outcome. However, if both prisoners choose to betray, they will end up with a worse outcome than if they both had cooperated\cite{king1988prisoner, gomez2011evolutionary, stewart2012extortion}.

In the 1980s, Robert Axelrod organized two tournaments to study the performance of various strategies in the iterated Prisoner's Dilemma (IPD) and to determine which strategies could balance cooperation and betrayal\cite{axelrod1981evolution}. In IPD studies, strategies are often endowed with a degree of "memory," allowing individuals to recall outcomes of several previous rounds. It is generally believed that players with stronger memory perform better in repeated games. However, research indicates that long-term memory does not significantly advantage over short-term memory\cite{baek2016comparing}. As a result, memory-one strategies have become the most widely used framework in repeated games. Scholars have proposed several strategies within this framework, highlighting their benefits in specific contexts. In the aforementioned tournaments, a simple Tit-for-Tat (TFT) strategy won consecutively. The TFT strategy involves cooperating in the first round and then replicating the opponent's action from the previous round. This cooperative approach yielded excellent results by not initiating betrayal but responding to it, thus balancing cooperation and punishment\cite{nowak1992tit, dugatkin1991guppies}. Inspired by TFT, Robert Axelrod proposed the Generous-TFT (GTFT) strategy, which also starts with cooperation and continues if the opponent does. However, unlike TFT, GTFT forgives the opponent's betrayal with a certain probability\cite{wedekind1996human}. This strategy maintains TFT's ability to establish cooperation while adding tolerance, helping to avoid vicious cycles and promoting long-term cooperation. In the 1950s, psychologist Donald Hebb introduced the concept of Win-Stay, Lose-Shift (WSLS). In the 1990s, Nowak and Sigmund formally defined this strategy. Its principle is simple: if the previous round's result was favorable, maintain the same decision in the current round; otherwise, change the decision\cite{nowak1993strategy, imhof2007tit}. In 2012, Press and Dyson introduced the Zero-Determinant (ZD) strategy, which can unilaterally control the opponent's payoff and enforce a linear relationship between their payoffs. Unlike the aforementioned strategies with clear rules, the ZD strategy encompasses a cluster of strategies based on repeated games\cite{press2012iterated}. Notably, its payoff is the expected long-term payoff rather than the exact payoff in any specific round.

Complex networks are systems composed of numerous interconnected nodes, which can be individuals, organizations, or other social units in reality. In these networks, the decisions and behaviors of individuals are often influenced by their surrounding nodes, leading to complex interactions and dynamic evolution\cite{scata2016combining, nowak1992evolutionary, ariful2018influence}. Studying evolutionary games on complex networks allows us to better understand the dynamics of interactions, cooperation, and competition among individuals, offering solutions to real-world social problems.

Game theory typically assumes that participants are completely rational and symmetric. However, in reality, participants often differ in identity, characters and assets, which significantly influence their decisions and payoffs\cite{du2009effect}. Introducing these differences makes game models more realistic and capable of accurately reflecting the complex interactions in the real world\cite{cuesta2015reputation, jian2021impact, chen2009diversity, ye2017evolutionary}. These participants have varying goals and resources in the game, leading to different strategies. For instance, resource-rich participants may be more willing to take risks, while resource-limited participants might prefer conservative strategies. By considering differences in varied attributes, more complex and optimized game models can be designed, resulting in fairer and more effective solutions. 

The main structure of this paper is as follows: Chapter 1 is a brief introduction to game theory, the prisoner's dilemma, complex networks, and asymmetric games. Chapter 2 describes the models and experimental procedures used in the study in detail. Chapter 3 presents and analyzes the experimental results. Chapter 4 summarizes the work conducted in the paper.

\section{Models and Settings}
\subsection{The Prisoner's Dilemma}
The Prisoner’s Dilemma (PD) is a classic game theory model. In the traditional PD, there are two participants, \textbf{X} and \textbf{Y}, each with the same options: cooperate (C) or defect (D). Let the cost of cooperation be denoted as $c$, and the benefit obtained be denoted as $b$\cite{lotfi2022effect, zhang2022reputation}. In a single round, if both \textbf{X} and \textbf{Y} choose to cooperate, they both receive the same payoff $R(reward)=b-c$. If \textbf{X} chooses to cooperate while \textbf{Y} chooses to defect, the naive cooperator \textbf{X} incurs the cost of cooperation, resulting in payoff $S(sucker)=-c$, while the greedy defector \textbf{Y} avoids the cooperation cost and directly gains payoff $T(temptation)=b$. If both \textbf{X} and \textbf{Y} choose to defect, neither incurs the cost, and neither gains the benefit, resulting in a payoff $P(punish)=0$ for both. Generally, it holds that $b>c>0$ and $T>R>P>S$\cite{hilbe2013evolution, bi2023heterogeneity}. In this paper, we set $b=4$ and $c=1$, yielding the following payoff matrix.

\begin{table}[h]
\centering
\captionsetup{justification=centering}
\begin{tabular}{|c|>{\centering\arraybackslash}m{1.5cm}|>{\centering\arraybackslash}m{1.5cm}|}
\hline
 & \textbf{C} & \textbf{D} \\ \hline
\textbf{C} & R(3) & S(-1) \\ \hline
\textbf{D} & T(4) & P(0) \\ \hline
\end{tabular}
\vspace{0.5cm}
\caption{Payoff Matrix}
\label{tab1}
\end{table}

\subsection{Memory-one Strategy}
In this study, all individuals are assumed to adopt a memory-one strategy $\textbf{p}=(p_R, p_S, p_T, p_P)$ for their interactions. The four parameters in this model correspond to the probability that an individual will choose to cooperate in the current round, based on the outcomes of the previous round being $\textbf{XY}=CC$, $\textbf{XY}=CD$, $\textbf{XY}=DC$ and $\textbf{XY}=DD$ respectively. Besides, these probabilities satisfy the condition $p_R \in [0, 1]$, $p_S \in [0, 1]$, $p_T \in [0, 1]$ and $p_P \in [0, 1]$ \cite{ichinose2018zero}. Specifically, to distinguish the strategies of both parties in the game, when \textbf{X} and \textbf{Y} engage in a repeated PD, the strategy of individual \textbf{X} is denoted as $\textbf{p}=(p_1, p_2, p_3, p_4)$, where $p_1$, $p_2$, $p_3$ and $p_4$ represent the probabilities of \textbf{X} choosing to cooperate given the previous round's outcomes of $\textbf{XY}=CC$, $\textbf{XY}=CD$, $\textbf{XY}=DC$ and $\textbf{XY}=DD$ respectively. Similarly, the strategy of \textbf{Y} is denoted as $\textbf{q}=(q_1, q_2, q_3, q_4)$, where $q_n$ represent the probabilities of \textbf{Y} choosing to cooperate given the previous round's outcomes of $\textbf{XY}=CC$, $\textbf{XY}=DC$, $\textbf{XY}=CD$ and $\textbf{XY}=DD$ respectively.

\subsection{Asymmetric Element}
The asymmetry in this study is reflected in the concept of "wealth value". Wealth value integrates factors such as reputation, status and capital, leading to varying returns for individuals in the game. In this study, the wealth value $k$ ranges from $(0, 10)$ to reflect the differences between the individuals\cite{han2023complex}. 

Assuming the total number of individuals in the group is $N$, and $N$ random numbers within the range $(0, 10)$ are generated and assigned to each individual as their initial wealth value before the first round.

From Table 1, the basic payoff matrix under symmetric games can be derived as follows.

\begin{equation}
    A=
    \begin{bmatrix}
        R & S \\
        T & P
    \end{bmatrix}
    =
    \begin{bmatrix}
        b-c & -c \\
        b & 0
    \end{bmatrix}
    =
    \begin{bmatrix}
        3 & -1 \\
        4 & 0
    \end{bmatrix}
\end{equation}
For \textbf{X} and \textbf{Y}, with wealth values $k_1$ and $k_2$ respectively, the payoff matrix is defined as follows.

\begin{equation}
    A_\textbf{X}=
    \begin{bmatrix}
        R & S \\
        T & P
    \end{bmatrix}
    =
    \begin{bmatrix}
        k_1(b-c) & -k_1c \\
        k_1b & 0
    \end{bmatrix}
    =
    \begin{bmatrix}
        3k_1 & -k_1 \\
        4k_1 & 0
    \end{bmatrix}
\end{equation}

\begin{equation}
    A_\textbf{Y}=
    \begin{bmatrix}
        R & S \\
        T & P
    \end{bmatrix}
    =
    \begin{bmatrix}
        k_2(b-c) & -k_2c \\
        k_2b & 0
    \end{bmatrix}
    =
    \begin{bmatrix}
        3k_2 & -k_2 \\
        4k_2 & 0
    \end{bmatrix}
\end{equation}

This settings integrate the impact of wealth value into the payoff matrix. It can be understood as follows: due to the differing identities, statuses, and assets of the two participants in the game, they can only obtain returns that match their positions. For instance, in a special PD, two prisoners have committed a crime together, but their sentences differ due to their different roles in the crime. Suppose prisoner \textbf{X}'s crime is more severe, leading to a longer sentence, while prisoner \textbf{Y}'s crime is less severe, resulting in a shorter sentence. If they both cooperate, they will achieve outcomes proportional to their sentences: 2 years and 1 year respectively. If \textbf{X} cooperates and \textbf{Y} defects, \textbf{Y} will be immediate released, while \textbf{X} will stay in the prison for 10 years. If on contrast, \textbf{X} will get immediate release, while \textbf{Y} will receive 8 years. If they both defect, they will receive relatively bad outcomes: 5 years and 3 years respectively.

\subsection{Payoff Calculation}
In a symmetric game, where the wealth values of \textbf{X} and \textbf{Y} are both 1, the payoff vector for \textbf{X} is defined as $R_X=(3, -1, 4, 0)$, and correspondingly, the payoff vector for \textbf{Y} is defined as $R_Y=(3, 4, -1, 0)$. For the asymmetric game, due to changes in the payoff matrix, the payoff vectors for both players become $R_\textbf{X}=(3k_1, -k_1, 4k_1, 0)$ and $R_\textbf{Y}=(3k_2, 4k_2, -k_2, 0)$ respectively. For both scenarios, after a single interaction, the expected payoffs for \textbf{X} and \textbf{Y} can be calculated using the following formula.

\begin{equation}
\begin{aligned}
    r_\textbf{X}=\frac{\mu \cdot \mathbf{R}_X}{\mu \cdot 1}=\frac{D(p, q, R_\textbf{X})}{D(p, q, 1)}
    \\
    r_\textbf{Y}=\frac{\mu \cdot \mathbf{R}_Y}{\mu \cdot 1}=\frac{D(p, q, R_\textbf{Y})}{D(p, q, 1)}
\end{aligned}
\end{equation}

Among them, $\mu$ is the stationary vector of matrices $p$ and $q$. In addition,

\begin{equation}
\begin{aligned}
    \mu \cdot h &= D(p, q, h) = \left|
    \begin{array}{cccc}
    -1+p_1 q_1 & -1+p_1 & -1+q_1 & h_1 \\
    p_2 q_3 & -1+p_2 & q_3 & h_2 \\
    p_3 q_2 & p_3 & -1+q_2 & h_3 \\
    p_4 q_4 & p_4 & q_4 & h_4
    \end{array} 
    \right| \\
    \mathbf{h} &= \begin{bmatrix}
    h_1 \\
    h_2 \\
    h_3 \\
    h_4
    \end{bmatrix}
\end{aligned}
\end{equation}

In this paper, the expected payoff for each node in the network is defined as the average payoffs obtained by itself from interactions with all its neighbors.

\subsection{Strategy Updating Method}
In complex networks, the actions of individuals can be divided into interaction and updating. During the simulation process, interaction involves each individual playing a two-person asymmetric PD game with all its neighbors and obtaining the corresponding payoff. Updating occurs after each individual completes a round of games: each individual randomly selects one of its neighbors to compare payoffs and decide whether to update their strategy.

For the basis of deciding whether to update the strategy, we choose to use the Fermi function in this paper. For \textbf{X}, whose neighbor set is $\textbf{P}$, \textbf{X} obtains an average payoff $r_\textbf{X}$ from the recent round with all players in $\textbf{P}$. At this point, \textbf{X} randomly selects a player \textbf{Y} from $\textbf{P}$, who has obtained an average payoff $r_\textbf{Y}$ in the same round. According to the Fermi dynamics, $r_\textbf{X}$ will adopt $r_\textbf{Y}$'s strategy in the next round with a probability given by $w$ or will continue using its current strategy with a probability of $1-w$\cite{he2020reputation, szabo1998evolutionary}.

\begin{equation}
    w=\frac{1}{1+\exp{\frac{r_\textbf{X}-r_\textbf{Y}}{k}}}
\end{equation}

In the denominator of the formula, $k$ represents the rationality level of individuals in the network. As $k$ approaches infinity, individuals gradually tend to make completely random choices regarding whether to update their strategy. Conversely, as $k$ approaches zero, individuals become fully rational, meaning they will adopt the other individual's strategy as long as the other's expected payoff is higher than their own.

According to previous research, in symmetric games, $k$ is often set to 1. However, this value cannot be directly applied to asymmetric games. For example, consider a game between \textbf{X} with huge wealth and \textbf{Y} with relatively low wealth. Because \textbf{X} has a much larger principal, he can obtain significantly higher payoffs compared to \textbf{Y}. However, \textbf{Y} should not easily adopt \textbf{X}'s strategy, because with his relatively smaller principal, adopting \textbf{X}'s strategy will not lead to a significant increase in his payoff. In this paper, the parameter $k$ is set to 8 to match the outcomes of symmetric games where $k=1$.

\begin{figure}[H]
    \begin{center}
	\includegraphics[width=0.98\textwidth]{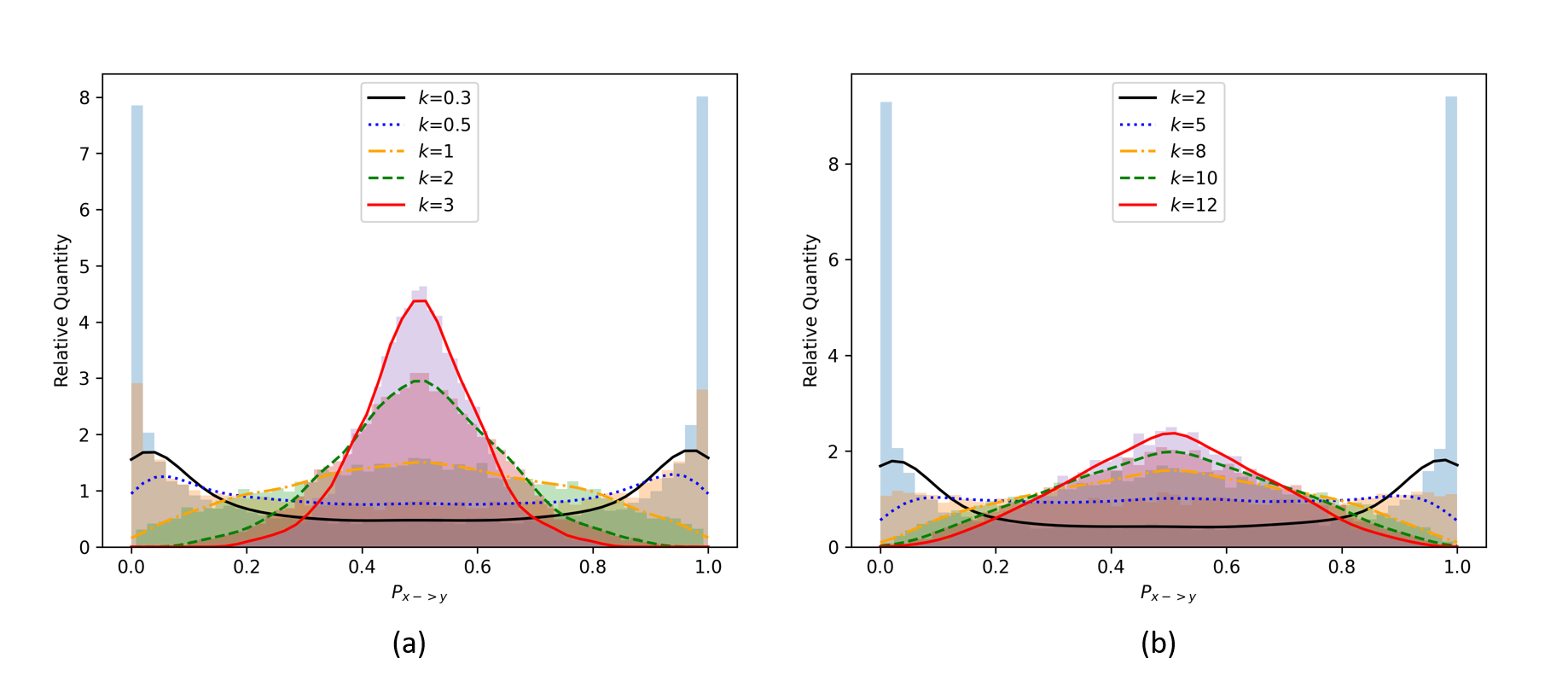}
	\caption{(a). Distribution Map of Conversion Probability on Symmetric Network.   (b). Distribution Map of Conversion Probability on Asymmetric Network.}
	\label{fig1}
    \end{center}
\end{figure}

\subsection{Wealth Updating Method}
In this paper, all wealth values are defined within the interval $(0, 10)$. Initially, each "participant" in the network randomly receives a wealth value within this range. During one round, participants receive an average payoff, which depends on their original wealth value and strategy choice. Given the bounded interval for wealth values, the average payoff will also fall within a specific range. After this single round, the original wealth of all participants and their average payoff from that round are summed. This total is then normalized to the interval $(0, 10)$ to ensure a unified standard for wealth values, preventing any strong individual's wealth from growing excessively and causing the strategy set to converge too quickly. Additionally, it is important to emphasize that after each round of wealth updates, the payoff vector $R=(3k, -k, 4k, 0)$ of each individual will also change accordingly. This means the strategy dynamics are continually influenced by the updated wealth and payoff values, maintaining a dynamic and adaptive system throughout the simulation process.

\section{Results}

This chapter is divided into three sections. The first section provides a brief classification and discussion of the strategy domain. The second section elaborates on the evolutionary game of asymmetric Prisoner's Dilemma on BA scale-free network and analyzes the result. The third section conducts supplementary experiments on the evolutionary outcomes.

\subsection{Classification and Discussion of Strategy Domain}
\subsubsection{Analysis of Win Rate Curves at Different Cooperation Levels}

An analysis is conducted to understand the impact of each component of $S$ on the payoff (win rate) against random strategies. To perform it, for each $p \in [0, 1$ in $S=(p_1, p_2, p_3, p_4)$, we choose $p=0.2$, $p=0.5$ and $p=0.8$ to represent "low cooperation willingness", "medium cooperation willingness" and "high cooperation willingness" respectively. $p_1$, $p_2$, $p_3$ and $p_4$ are controlled separately for simulations and statistical classification. 10,000 random strategies are generated to calculate the win rate of $S$ against these random strategies. Partial experimental results are presented below, and others are presented appendix.

\begin{figure}[H]
\makebox[\textwidth][c]{\includegraphics[width=1\textwidth]{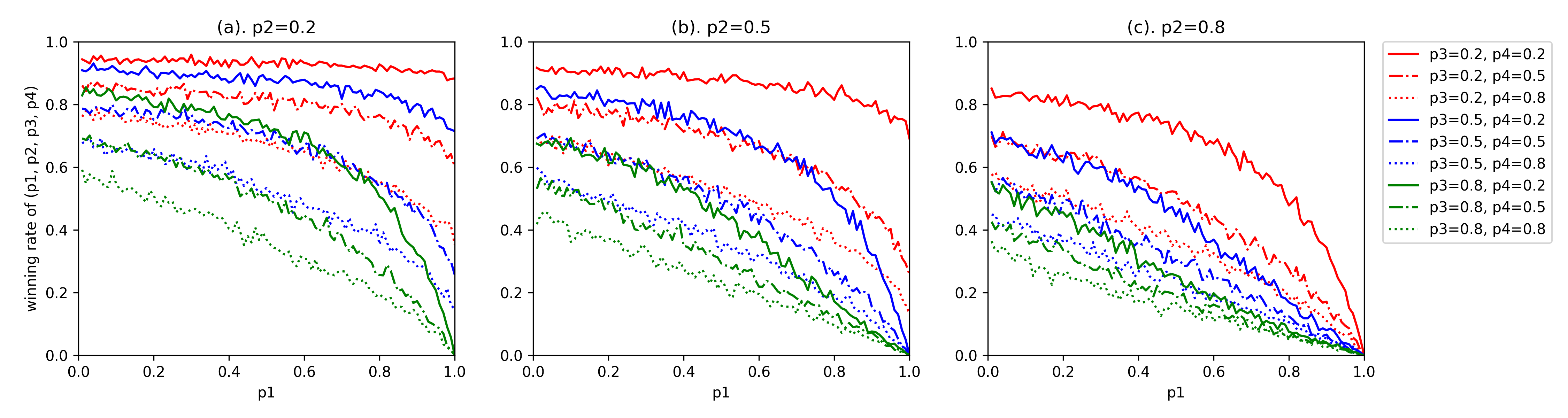}}
\caption{$p_1$'s Impact on Win Rate against Random Strategies When $p_2=0.2$, $p_2=0.5$ and $p_2=0.8$ Respectively}
\label{fig2}
\end{figure}

In Figure \ref{fig2}, the curves indicates that the performance of strategy $S$ against random strategies is negatively impacted by increases in the values of parameters $p_1$, $p_2$, $p_3$ and $p_4$.

Specifically, the trends shown by the solid, dashed, and dotted curves of the same color suggest that increase in parameter $p_4$ has a detrimental effect on the win rate. Similarly, the trends shown by the same line types in different colors indicate that increase in parameter $p_3$ performs the same. Furthermore, the overall downward trends observed across the three figures as the parameter $p_1$ increase, as well as the progressive reductions in the values of the corresponding curves, demonstrate that growth in both parameters $p_1$ and $p_2$ adversely affect the win rate.

\subsubsection{Comparison of Each $p$}

To analyze the relative impact of the four components on win rates, heatmaps were generated under different combinations of these components. For instance, when comparing the relative effects of $p_1$ and $p_1$, fixed values were assigned to $p_3$ and $p_4$. As $p_1$ and $p_2$ varied from low to high, 1000 random strategies were generated at each sampling point, and the win rate against these random strategies was computed to create a heatmap of the distribution. Below is a selection of the experimental results, with the remaining results displayed in appendix.

\begin{figure}[H]
    \captionsetup{justification=raggedright,singlelinecheck=false}
	\makebox[\textwidth][c]{\includegraphics[width=1.\textwidth]{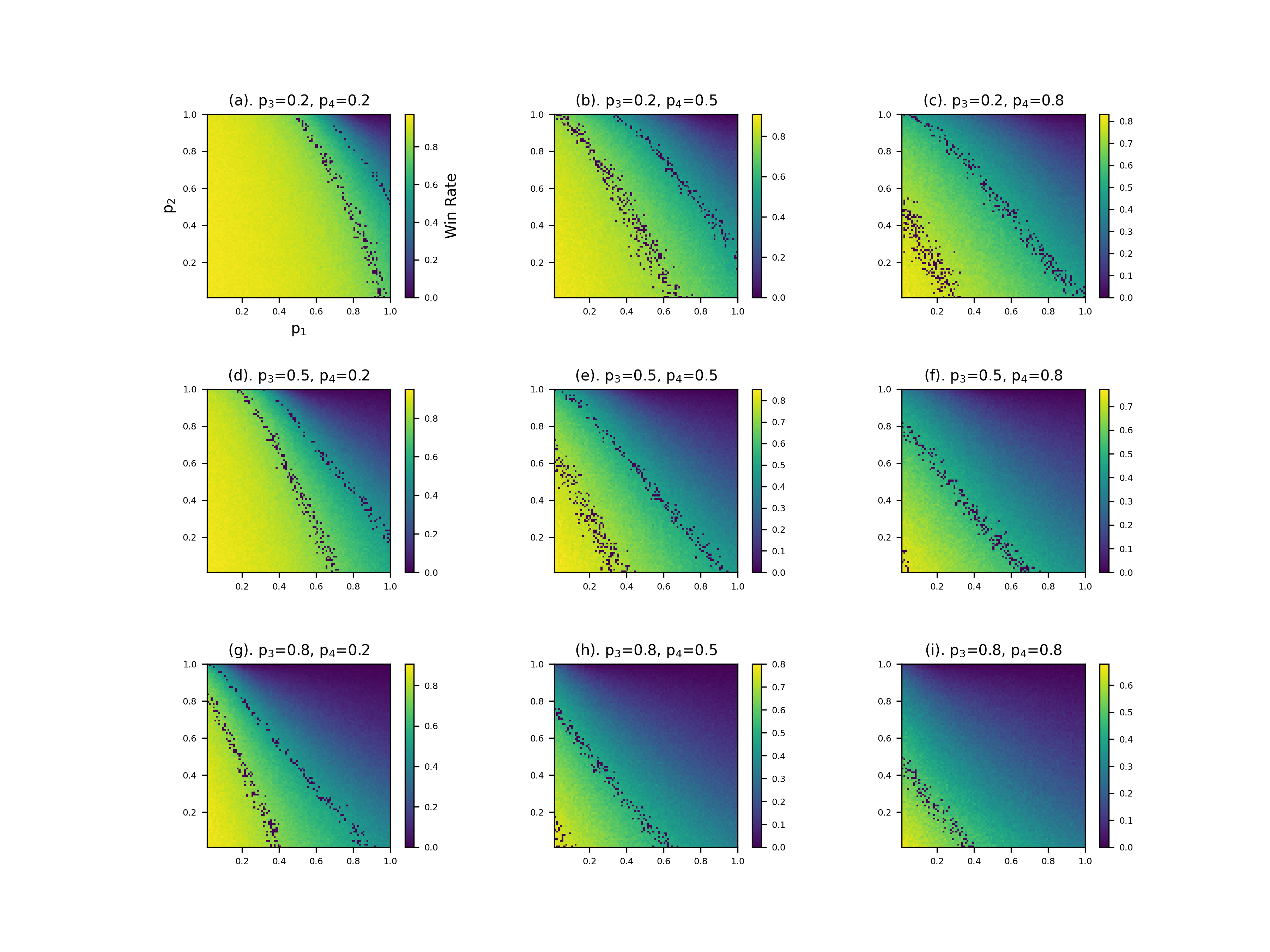}}
	\captionof{figure}{Impact of $p_1$ and $p_2$ on Win Rate When $p_3$ and $p_4$ Diverse}
	\label{fig3}
\end{figure}

Figure \ref{fig3} illustrates the impact of $p_1$ and $p_2$ on the win rate of $S=(p_1, p_2, p_3, p_4)$ against random strategies when $p_3$ and $p_4$ are set to 0.2, 0.5 and 0.8 respectively. The horizontal and vertical axes have consistent meanings, and the two distinct color spots in the figure represent win rates ($W$) satisfying $0.495\le W \le0.505$ and $0.745\le W \le0.755$ (with only the former appearing in Figure (i)). Taking Figure (a) as an example, both $p_3$ and $p_4$ are at relatively low levels (corresponding to the practical scenario where the current round's probability of cooperation is low after the previous round's defection by the player who uses strategy $S$). To improve the player's win rate, $p_1$ and $p_2$ should be maintained at low levels, which is consistent with the preliminary conclusions obtained earlier. Furthermore, maintaining $p_2$ at a low level is more conducive to achieving better results than reducing $p_1$, indicating that "reducing $p_1$" is more beneficial for victory compared to "reducing $p_2$". Given the practical significance of $p_1$ and $p_2$, it can be inferred that "greedily" defecting can reap greater benefits when both sides cooperated in the previous round, while showing some tolerance when the S-user cooperated, and the opponent defected in the previous round, might also yield favorable outcomes.

\begin{figure}[H]
    \captionsetup{justification=raggedright,singlelinecheck=false}
	\makebox[\textwidth][c]{\includegraphics[width=1\textwidth]{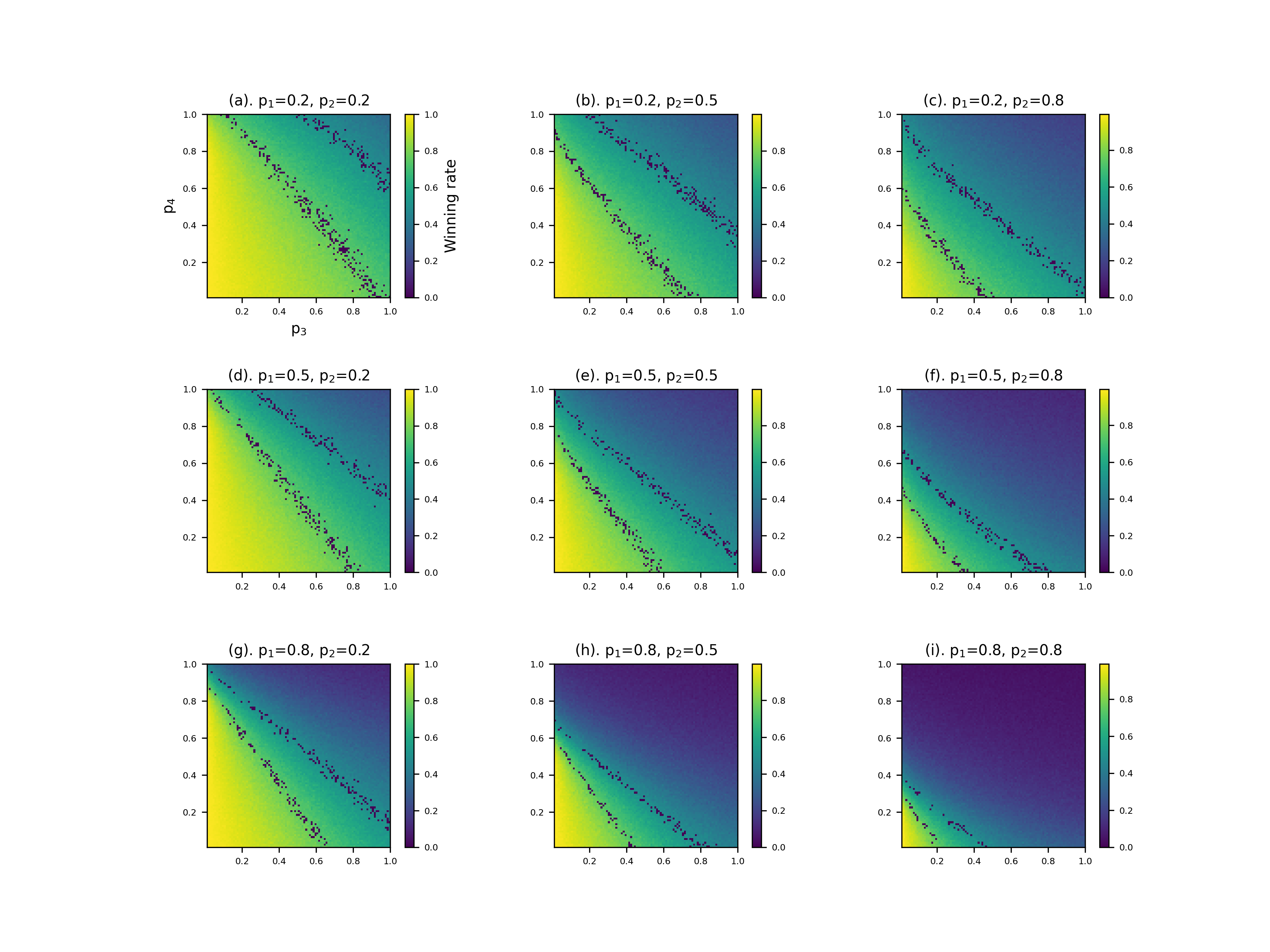}}
	\captionof{figure}{Impact of $p_3$ and $p_4$ on Win Rate When $p_1$ and $p_2$ Diverse}
	\label{fig4}
\end{figure}

Figure \ref{fig4} illustrates the impact of $p_3$ and $p_4$ on the win rate of $S=(p_1, p_2, p_3, p_4)$ against random strategies when $p_1$ and $p_2$ are set to 0.2, 0.5 and 0.8 respectively. In this example, $p_1$ is at a high level and $p_2$ is at a moderate level. This corresponds to a situation where the player has a high probability of cooperating if both players cooperated in the previous round, and a moderate probability of cooperating if the player cooperated but the opponent defected in the previous round. The results show that to achieve a high win rate, $p_3$ and $p_4$ should be maintained at relatively low levels. Specifically, keeping $p_3$ at a low level is more beneficial for outperforming the random strategy than keeping $p_4$ at a low level. This suggests that when both players defected in the previous round, it may be advantageous to "reconcile" to some degree, rather than continuing to defect. Conversely, when the player cooperated but the opponent defected in the previous round, the player should consider continuing to "exploit" the opponent's goodwill, as this can lead to more favorable outcomes.

Our research on the four-parameter set reveals that the parameters have varying degrees of impact on the win rate, with $p_1$ being the most influential, followed by $p_2$, $p_3$ and $p_4$. This suggests that if the strategy $S=(p_1, p_2, p_3, p_4)$ has to focus on improving one parameter, it would be most beneficial to prioritize keeping $p_1$ at a relatively low level, while considering a moderate increase in $p_4$. Interpreting this in practical terms, when facing the outcome of mutual cooperation in the previous round, it would be the optimal choice for strategy $S$ to lean towards defection. Similarly, when confronting the outcome of mutual defection in the previous round, continuing to defect would certainly be the best option. However, moderately increasing the probability of cooperation in this scenario can help maintain one's own payoff while appearing less purely self-interested.

\subsection{Asymmetric IPD on BA Scale-Free Network}
\subsubsection{Experimental Settings and Steps}
The BA scale-free network model is a dynamic network model commonly used to generate scale-free networks. It simulates the "rich-get-richer" phenomenon observed in social networks, where nodes with higher degrees are more likely to attract new connections\cite{mao2017fast}. In our experiment, the network is defined with a final node count of $n=1000$, and each newly added node has an initial degree $m=20$. This results in a total of $e=19600$ edges.

In the experiment, each node in the network represents an individual with an initial state characterized by a randomly assigned memory-one strategy. These individuals engage in interactions with all their neighbors in each round of the game, generating payoffs based on these interactions. After each round, individuals update their strategies based on the payoffs they and their neighbors received. This iterative process continues until the network reaches an equilibrium state. An equilibrium state is defined as either a state where only one strategy remains across the network, or a dynamic equilibrium where, after 2000 rounds of games, several (usually no more than three) strategies persist.

\begin{figure}[!htbp]
    \centering
    \captionsetup{justification=raggedright,singlelinecheck=false}
    \includegraphics[width=2in]{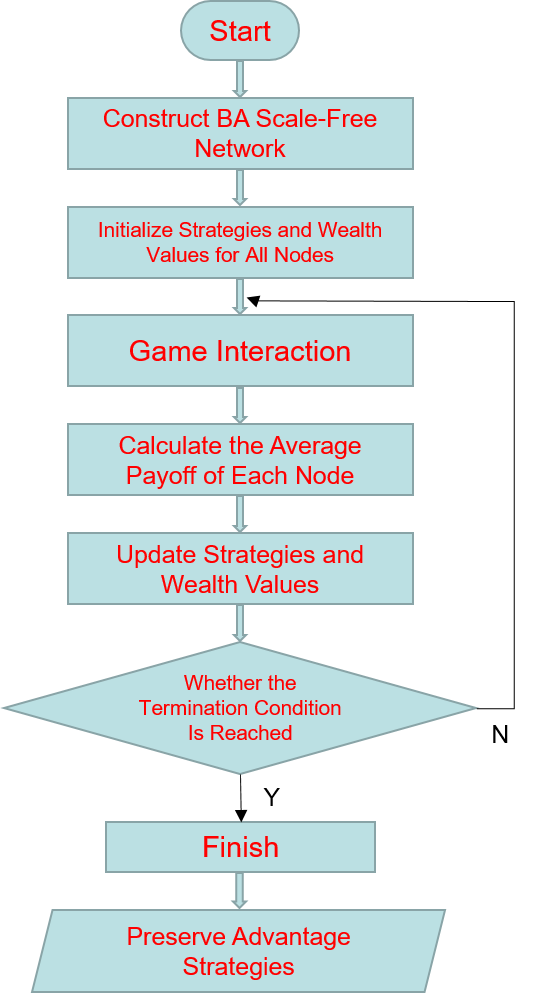}
    \caption{Experimental Flowchart}
    \label{fig5}
\end{figure}

\subsubsection{Experimental Results}
Following the aforementioned method, 1000 repeated experiments are conducted, resulting in over 1200 distinct dominant strategies. The strategies were categorized and analyzed using clustering algorithms.

Due to the narrow distribution of the strategy set within the interval $p \in [0, 1]$, clustering algorithms such as DBSCAN, which require specifying point spacing and neighborhood size, presented significant challenges. Therefore, we utilized the K-Means algorithm, supported by the elbow method and silhouette coefficient and discover that the appropriate number of clusters is 6. Table \ref{tab2} presents the coordinates of the center points and the number of individuals in each cluster.

\begin{table}[H]
\renewcommand{\arraystretch}{1}
\setlength{\arrayrulewidth}{0.5mm}
\setlength{\tabcolsep}{5pt}
\captionsetup{justification=raggedright,singlelinecheck=false}
\centering
\begin{tabular}{
    |>{\centering\arraybackslash}m{1.5cm}  
    |>{\centering\arraybackslash}m{7cm}   
    |>{\centering\arraybackslash}m{1.5cm}|  
    }
    \hline
    Number & Center Points of Clusters & Cluster Sizes \\
    \hline
    1 & $[0.2761, 0.6548, 0.1609, 0.2059]$ & 229 \\
    \hline
    2 & $[0.7709, 0.3186, 0.1537, 0.1782]$ & 256 \\
    \hline
    3 & $[0.1964, 0.3716, 0.6072, 0.6647]$ & 170 \\
    \hline
    4 & $[0.7031, 0.3372, 0.7134, 0.3185]$ & 155 \\
    \hline
    5 & $[0.6894, 0.3174, 0.1914, 0.7038]$ & 167 \\
    \hline
    6 & $[0.2365, 0.1649, 0.1814, 0.2794]$ & 284 \\
    \hline
\end{tabular}
\vspace{0.5cm}
\caption{Clusters' Information}
\label{tab2}
\end{table}

To further investigate the characteristics of each cluster, we recorded the payoffs when clusters confronted each other, as well as their win rates and average payoffs against 10,000 random strategies. The results are shown in Table 3.

\begin{table}[h]
\renewcommand{\arraystretch}{1}
    \centering
    \captionsetup{justification=raggedright,singlelinecheck=false}
    \setlength{\tabcolsep}{5pt}
    \begin{tabular}{
        |>{\centering\arraybackslash}m{1.0cm}  
        |>{\centering\arraybackslash}m{1.5cm}  
        |>{\centering\arraybackslash}m{1.5cm}  
        |>{\centering\arraybackslash}m{1.5cm}  
        |>{\centering\arraybackslash}m{1.5cm}  
        |>{\centering\arraybackslash}m{1.5cm}  
        |>{\centering\arraybackslash}m{1.5cm}|
    }
        \hline
         \rule{0pt}{1.5em} & \textbf{$S_1$} & \textbf{$S_2$} & \textbf{$S_3$} & \textbf{$S_4$} & \textbf{$S_5$} & \textbf{$S_6$} \\
        \hline
        \textbf{$S_1$} & \makecell{0.9122} & \makecell{\textcolor{red}{0.6274}} & \makecell{1.7089} & \cellcolor{green!25}\makecell{1.4411} & \makecell{1.6576} & \makecell{0.6205} \\
        \hline
        \textbf{$S_2$} & \makecell{\textcolor{red}{1.0206}} & \makecell{\textcolor{red}{0.7428}} & \makecell{\textcolor{red}{1.7172}} & \cellcolor{green!25}\makecell{\textcolor{red}{1.4502}} & \makecell{\textcolor{red}{1.7318}} & \makecell{\textcolor{red}{0.7242}} \\
        \hline
        \textbf{$S_3$} & \makecell{0.5270} & \makecell{\textcolor{red}{0.5229}} & \makecell{1.4123} & \cellcolor{green!25}\makecell{1.5617} & \makecell{1.4366} & \makecell{0.4060} \\
        \hline
        \textbf{$S_4$} & \cellcolor{green!25}\makecell{0.6135} & \cellcolor{green!25}\makecell{\textcolor{red}{0.6807}} & \cellcolor{green!25}\makecell{1.3813} & \cellcolor{green!25}\makecell{1.5878} & \cellcolor{green!25}\makecell{1.3996} & \cellcolor{green!25}\makecell{0.4989} \\
        \hline
        \textbf{$S_5$} & \makecell{0.5841} & \makecell{\textcolor{red}{0.6079}} & \makecell{1.4118} & \cellcolor{green!25}\makecell{1.5600} & \makecell{1.5476} & \makecell{0.3979} \\
        \hline
        \textbf{$S_6$} & \makecell{1.0201} & \makecell{\textcolor{red}{0.6828}} & \makecell{1.7492} & \cellcolor{green!25}\makecell{1.4266} & \makecell{1.7859} & \makecell{0.7163} \\
        \hline
    \end{tabular}
    \vspace{0.5cm}
    \caption{The Game Results of Each Cluster and the Win Rate and Average Payoff Facing Random Strategies}
    \label{tab3}
\end{table}

The table entries indicate the payoff obtained by the horizontal strategy when facing the vertical strategy. For example, the value 0.6274 in the cell corresponding to $S_1-S_2$ indicates that strategy $S_1$ gains a payoff of 0.6274 when confronting $S_2$.

The red entries show the payoffs of strategy  against all other strategies, highlighting that $S_2$ consistently achieves higher payoffs compared to its opponents, regardless of what strategy they choose. Additionally, $S_2$'s self-play payoff is lower than that of other strategies' self-play payoffs. This indicates that $S_2$ displays a significant advantage against random strategies, demonstrating a "self-bad, partner-worse" outcome in the asymmetric prisoner's dilemma, suggesting that such strategies can emerge and maintain a certain scale\cite{zhang2022self}.

The green-background entries represent the results of strategy $S_4$ against all other strategies. It is observed that $S_4$ always obtains a lower payoff compared to its opponents, who achieve relatively high payoffs. Moreover, $S_4$'s self-play payoff is higher than that of the other strategies' self-play payoffs, indicating its inclination towards seeking cooperation and ensuring better outcomes for both parties. In real life, this strategy corresponds to the "altruists" who prioritize the overall good. Furthermore, $S_4$ does not exhibit an advantage against random strategies and has the smallest number of individuals among the six strategy clusters, aligning with logical reasoning and common sense.

\subsection{The Evolution and Spread of Cooperative Strategies on Network}
In the previous section, we observed that in the evolutionary game of the asymmetric prisoner's dilemma on a BA scale-free network, the $S_4$ strategy might ultimately evolve. In real life, we always hope that groups tend towards cooperation. For example, parents teach their children in kindergarten to become good friends with their peers rather than encouraging hostility towards them. Similarly, in international affairs, powerful nations have always sought friendly exchanges with other nations to promote cooperation and mutual development. This part of the experiment aims to find a method to foster cooperation.

A crucial question is how to define the manifestation of enhanced cooperation. We propose the following research method: if the level of cooperation increases, the fitness of the population will improve, corresponding to an increase in the average payoff of the group in this experiment\cite{luo2016cooperation}. Based on this idea, we conducted two supplementary experiments.

In this section, we employ a BA scale-free network model with 100 nodes and an initial degree of 4 for each newly added node. Each node also has an initial "wealth value." Based on the experimental procedures described in the second section of this chapter, we made the following two modifications.

(1) Initial Entry of Strategy $S_4$

Before starting the experiment, we introduced strategy $S_4$ to the initial network according to the following rules.

Random Selection: Randomly select several nodes at the initial stage and assign them Strategy $S_4$.

Degree-Based Selection: Select several nodes at the initial stage based on their degree from high to low and assign them Strategy $S_4$.

Wealth-Based Selection: Select several nodes at the initial stage based on their initial wealth value from high to low and assign them Strategy $S_4$.

These operations aim to spread Strategy $S_4$ by leveraging the strategies of important nodes.

(2) Eliminating Low-Cooperation Strategies

During the strategy update phase after each round of the game, if a component $p$ of a node's strategy $S=(p_1, p_2, p_3, p_4)$ is lower than a certain threshold (indicating a very low level of cooperation), there is a certain probability that a neighboring node will be randomly selected (where all four components of the neighboring node's strategy $S'=(p_1', p_2', p_3', p_4')$ must be greater than this threshold) to adopt its strategy in the next round.

Each type of experiment described above was repeated 100 times, generating 100 dominant strategies that evolved. Each of these strategies was then subjected to 1000 tests against random strategies, and the average payoff of the random strategies was recorded. A distribution curve of these 100 average payoffs was plotted.

The understanding is that the initial entry of Strategy $S_4$ may influence the evolution of strategies within the network. If the influence is positive, the evolved dominant strategies should be able to promote cooperation within the group. And promoting group cooperation, in turn, is partly reflected in the increased fitness of the group when facing a random population, manifested as an increase in average payoff. The experiment yielded the following results.

\begin{figure}[H]
    \makebox[\textwidth][c]{\includegraphics[width=0.5\textwidth]{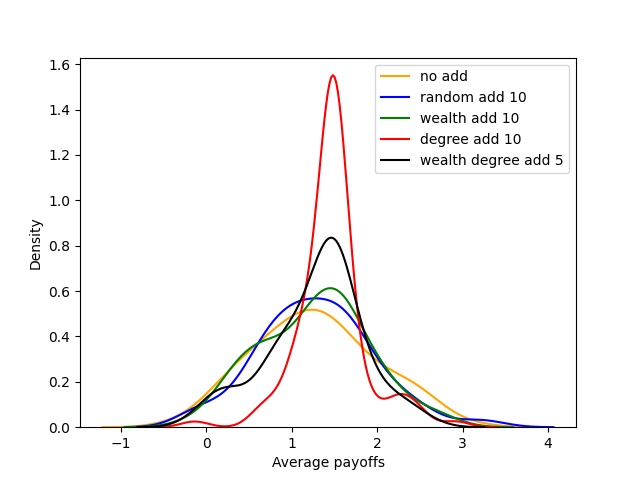}}
    \caption{The Impact of the Evolution Results on the Fitness of the Random Population after Adding Strategy $S_4$ in the First Round According to Different Rules}
    \label{fig6}
\end{figure}

From Figure \ref{fig6}, it can be observed that the evolution results of randomly introducing 10 strategies $S_4$ into the initial network did not have a significant impact on the fitness of the population when facing a random population. Similarly, selectively introducing 10 nodes with the highest wealth values, 10 nodes with the highest degrees, or a combination of 5 nodes with the highest wealth values and 5 nodes with the highest degrees to adopt strategy $S_4$ in the initial network did not significantly affect the average fitness of the population. This only resulted in the fitness of individuals in the population being more centered around an intermediate level.

\begin{figure}[H]
    \makebox[\textwidth][c]{\includegraphics[width=0.5\textwidth]{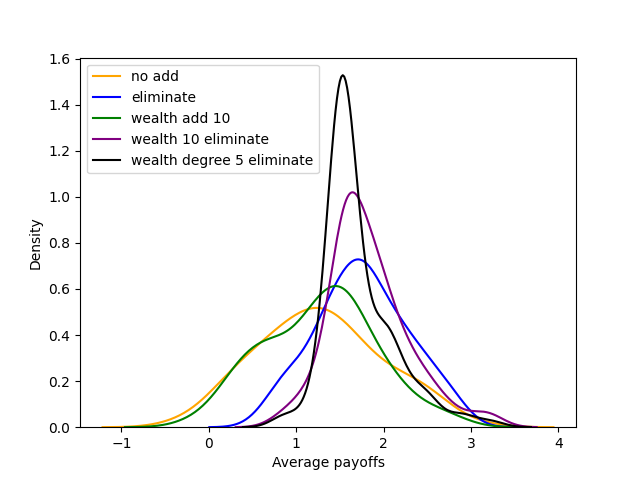}}
    \caption{The Impact of the Evolution Results on the Fitness of the Random Population after Adding Strategy $S_4$ in the First Round According to Different Rules and Adding Probabilistic Strategy Optimization Mechanism}
    \label{fig7}
\end{figure}

From Figure \ref{fig7}, it can be seen that selectively introducing 10 nodes with the highest wealth values to adopt strategy $S_4$ in the first round, combined with a 50\% probability of resetting low-cooperation strategies, resulted in a better distribution of population fitness. On one hand, this approach led to more individuals having intermediate fitness levels within the population. On the other hand, it also resulted in a certain number of high-fitness individuals. Furthermore, the range of the horizontal axis indicates that the mechanism of resetting low-cooperation strategies with a 50\% probability effectively eliminated individuals with negative payoffs in the random population. This suggests that the strategy reset mechanism can significantly enhance cooperation in the evolutionary outcome, thereby promoting cooperation during the evolution process.

\section{Conclusion and Discussion}
In the experiments above, the strategy domain of the framework used was classified and discussed, with a qualitative analysis of the impact of the four components of strategy  $S=(p_1, p_2, p_3, p_4)$ on the win rate of it against random strategies. To increase its win rate, the S-user should maintain a low level of cooperation. The experiments also compared the relative effects of the four components. In the study of $(p_1, p_2)$, it was found that "reducing $p_1$" is more conducive to victory compared to "reducing $p_2$". Specifically, in scenarios where both parties chose to cooperate in the previous round, a very low cooperation rate is optimal for the current round. Conversely, when the individual cooperated and the opponent defected in the previous round, a certain level of tolerance can help avoid mutually detrimental outcomes. In the study of $(p_3, p_4)$, it was found that "reducing $p_3$" is more conducive to victory compared to "reducing $p_4$".

In the strategy evolution experiments, the K-Means clustering algorithm identified six strategy clusters. Among these clusters, not only did a "self-bad, partner-worse" strategy cluster emerge, but a "altruists" strategy cluster also evolved. Similar to the zero-determinant strategies, the "self-bad, partner-worse" strategy can control the opponent's payoff to be lower than its own. The "altruists" strategy however, is at a disadvantage against other strategy clusters but achieves the highest payoff in self-play. The existence of this strategy is beneficial for the continuation and development of the group.

In the network evolution experiments involving the "altruists" strategy cluster, it was observed that introducing the "altruists" strategy cluster into the initial network according to different rules resulted in the evolved strategies generally placing the fitness of individuals in the random population at intermediate to high levels, with little impact on the average fitness. Additionally, after introducing a mechanism for eliminating low-cooperation strategies, the fitness of individuals in the population tended to be more centered around intermediate levels, and a certain number of high-fitness individuals also emerged. This mechanism overall enhanced the fitness of the population and proved to be a method for promoting cooperation.

The experiments provide a theoretical foundation for the evolutionary processes of social networks. Through the study of asymmetric prisoner's dilemma on weighted network, we uncover the relationships among strategies in complex evolutionary game environments, offering a framework for individuals and organizations to deploy effective countermeasures in practical decision-making. These findings not only enrich evolutionary game theory but also provide new perspectives and strategies for understanding and promoting cooperative behavior in social systems, thereby opening new avenues for enhancing overall population fitness and sustainable development.

\newpage
\appendix
\section{Other Win Rate Curves at Different Cooperation Levels}

\begin{figure}[H]
\makebox[\textwidth][c]{\includegraphics[width=1\textwidth]{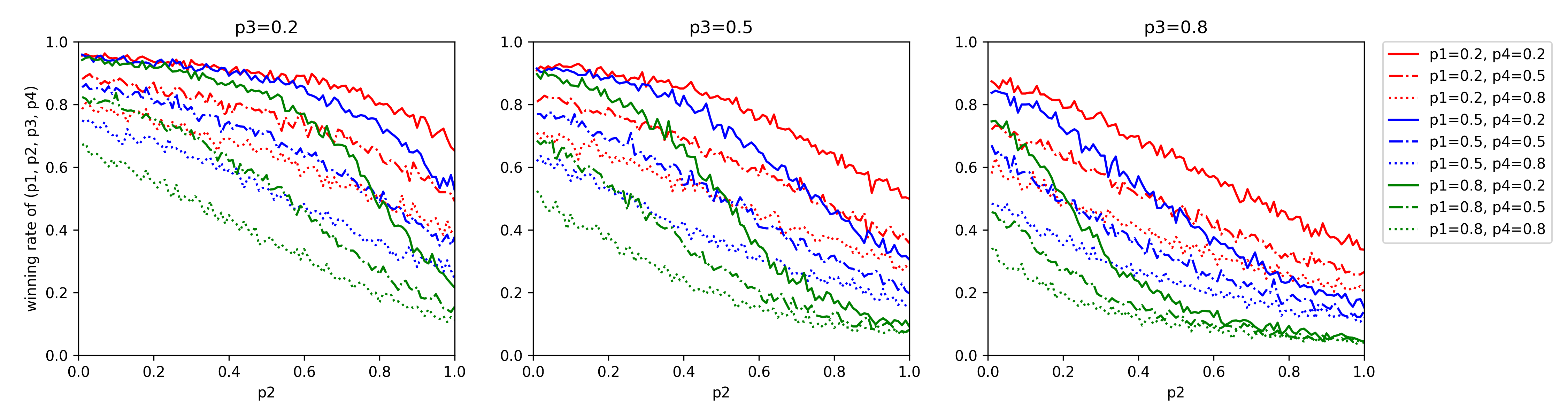}}
\caption{$p_2$'s Impact on Win Rate against Random Strategies When $p_3=0.2$, $p_3=0.5$ and $p_3=0.8$ Respectively}
\label{fig8}
\end{figure}

\begin{figure}[H]
\makebox[\textwidth][c]{\includegraphics[width=1\textwidth]{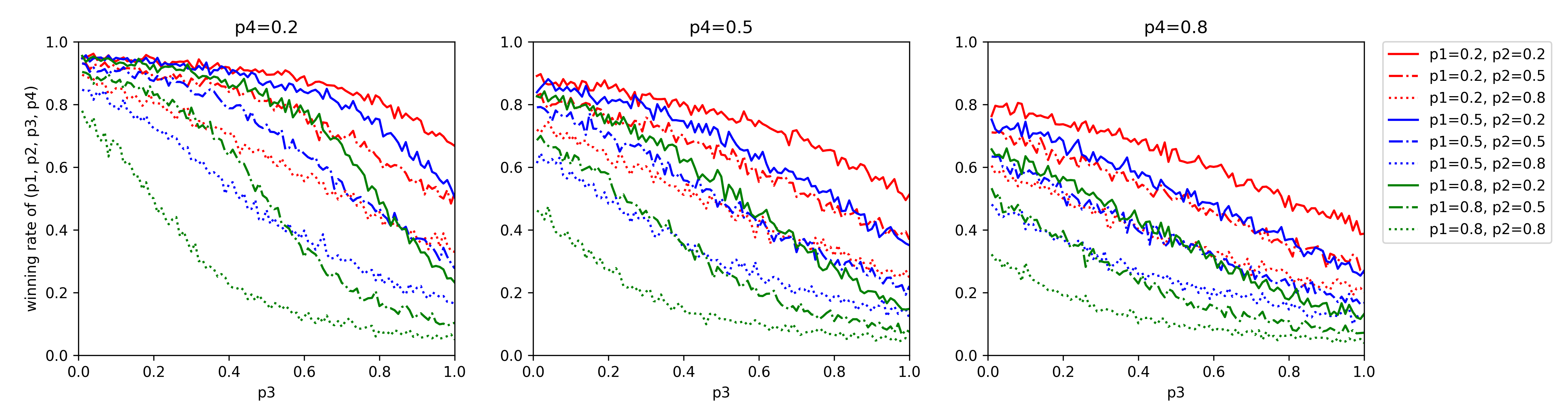}}
\caption{$p_3$'s Impact on Win Rate against Random Strategies When $p_4=0.2$, $p_4=0.5$ and $p_4=0.8$ Respectively}
\label{fig9}
\end{figure}

\begin{figure}[H]
\makebox[\textwidth][c]{\includegraphics[width=1\textwidth]{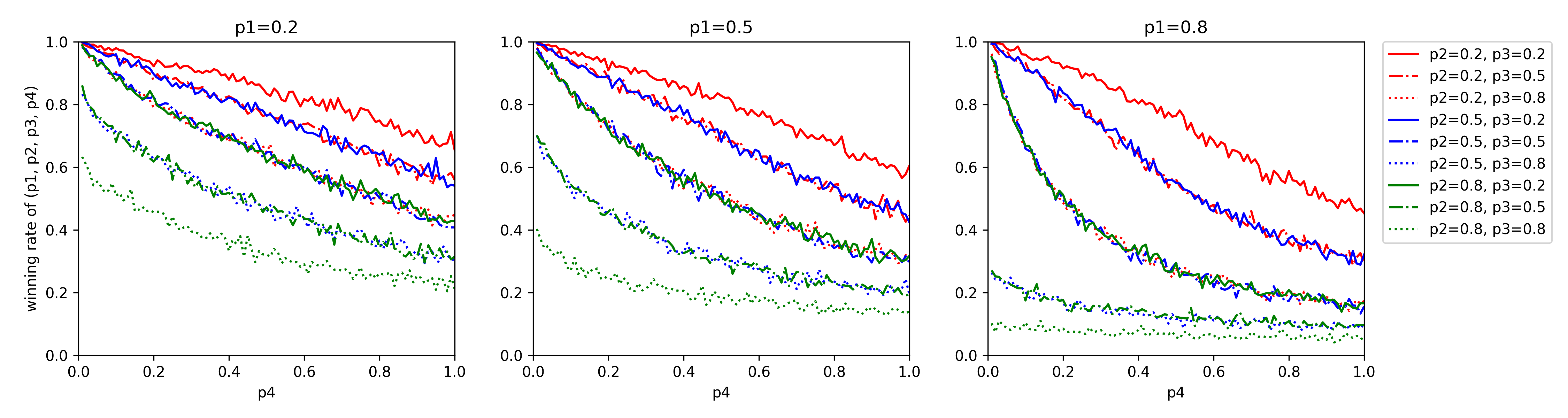}}
\caption{$p_4$'s Impact on Win Rate against Random Strategies When $p_1=0.2$, $p_1=0.5$ and $p_1=0.8$ Respectively}
\label{fig10}
\end{figure}

\section{Other Comparison of $p$}
\begin{figure}[H]
    \captionsetup{justification=raggedright,singlelinecheck=false}
	\makebox[\textwidth][c]{\includegraphics[width=1.2\textwidth]{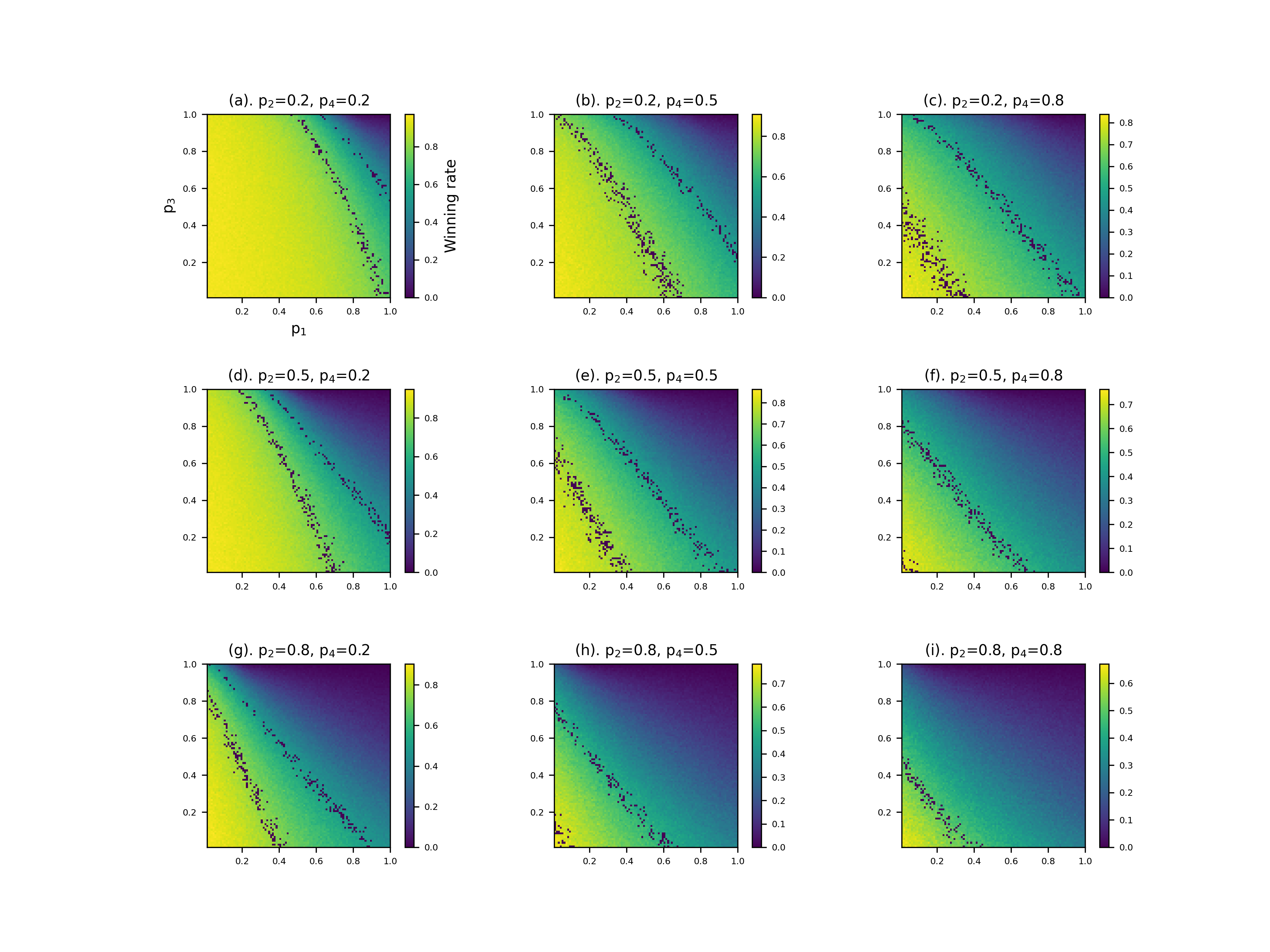}}
	\captionof{figure}{Impact of $p_1$ and $p_3$ on Win Rate When $p_2$ and $p_4$ Diverse}
	\label{fig11}
\end{figure}

\begin{figure}[H]
    \captionsetup{justification=raggedright,singlelinecheck=false}
	\makebox[\textwidth][c]{\includegraphics[width=1.2\textwidth]{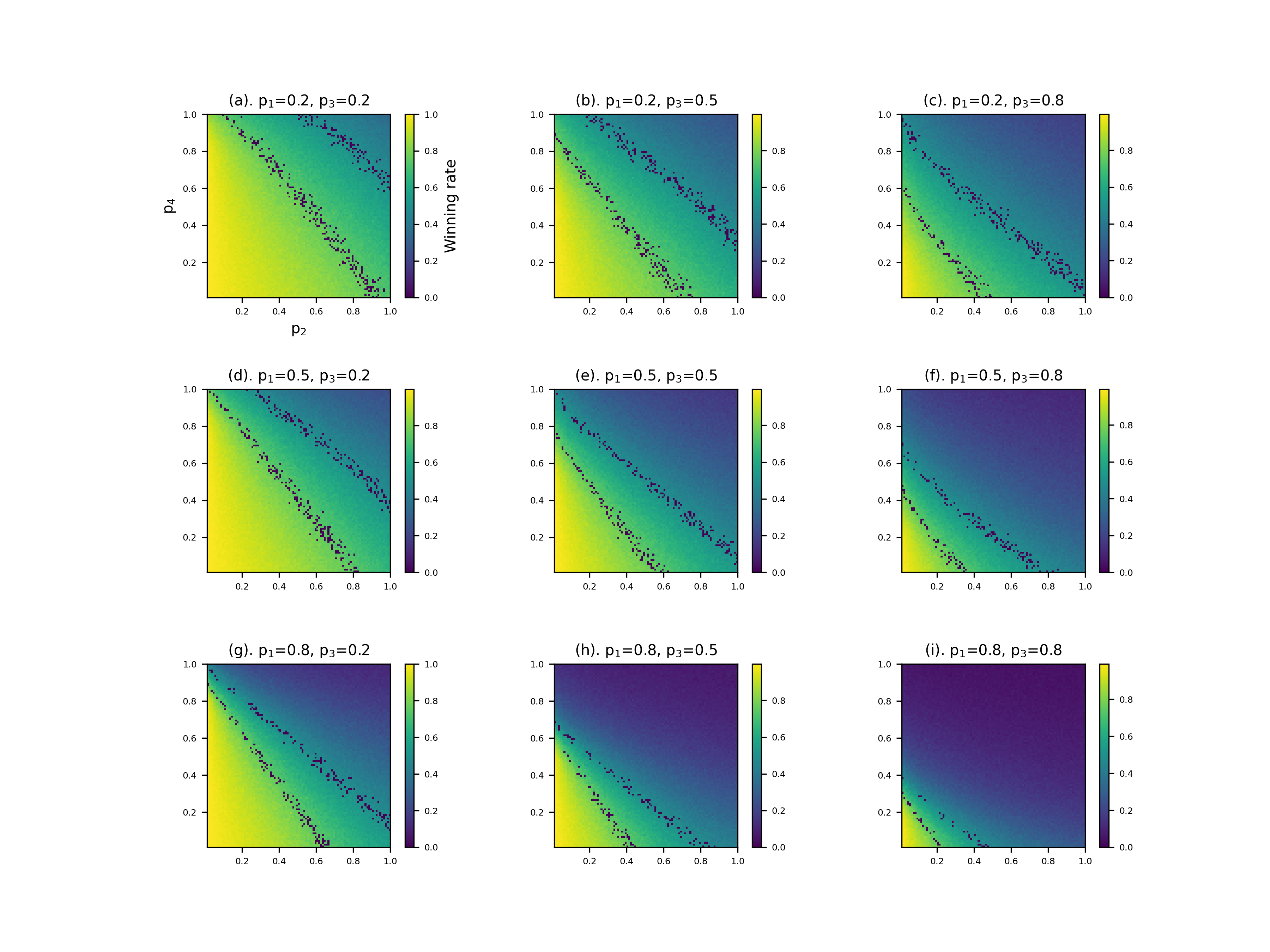}}
	\captionof{figure}{Impact of $p_2$ and $p_4$ on Win Rate When $p_1$ and $p_3$ Diverse}
	\label{fig12}
\end{figure}






\bibliographystyle{unsrt}
\bibliography{ref.bib}

\end{document}